\documentclass[superscriptaddress,twocolumn,showpacs,a4paper,
amssymb,amsmath,nobibnotes,aps,prd,
showkeys,
nofootinbib,notitlepage]{revtex4-1}
\usepackage{verbatim}
\usepackage[T1]{fontenc}
\usepackage[utf8]{inputenc}
\usepackage[american]{babel}
\usepackage{epsfig}
\usepackage{booktabs}
\usepackage{multirow}
\usepackage{dcolumn}
\usepackage{amsmath}
\usepackage{mathtools}
\usepackage{amsfonts}
\usepackage{amssymb}
\usepackage{ulem}
\usepackage{epstopdf}
\usepackage{bm}
\usepackage{siunitx}
\usepackage{braket}
\usepackage{enumitem}
\usepackage{soul}
\usepackage[table]{xcolor}
\usepackage{color}
\usepackage{transparent}
\usepackage{pifont}
\usepackage{enumitem}

\definecolor{navyblue}{rgb}{0.0, 0.0, 0.5}
\definecolor{royalblue}{rgb}{0.25, 0.41, 0.88}
\definecolor{cadmiumgreen}{rgb}{0.0, 0.42, 0.24}
\definecolor{blue-violet}{rgb}{0.54, 0.17, 0.89}
\definecolor{darkviolet}{rgb}{0.58, 0.0, 0.83}
\definecolor{orange(colorwheel)}{rgb}{1.0, 0.5, 0.0}

\usepackage{hyperref}
\hypersetup{
    colorlinks=true, 
    linkcolor=royalblue, 
    citecolor=magenta}

\usepackage{booktabs}
\usepackage{multirow}
\usepackage{dcolumn}
\usepackage{colortbl}

\begin{document}

\title{Constraining Scalar Charge in Neutron Stars through Gravitational Wave Signals from NS–BH Mergers}


\author{Rafael M. Santos}
\email{rafael.mancini@inpe.br}
\affiliation{Divis\~ao de Astrof\'isica, Instituto Nacional de Pesquisas Espaciais, Avenida dos Astronautas 1758, S\~ao Jos\'e dos Campos, 12227-010, SP, Brazil}

\author{Rafael C. Nunes}
\email{rafadcnunes@gmail.com}
\affiliation{Instituto de F\'{i}sica, Universidade Federal do Rio Grande do Sul, 91501-970 Porto Alegre RS, Brazil}
\affiliation{Divis\~ao de Astrof\'isica, Instituto Nacional de Pesquisas Espaciais, Avenida dos Astronautas 1758, S\~ao Jos\'e dos Campos, 12227-010, SP, Brazil}

\author{Jose C. N. de Araujo}
\email{jcarlos.dearaujo@inpe.br}
\affiliation{Divis\~ao de Astrof\'isica, Instituto Nacional de Pesquisas Espaciais, Avenida dos Astronautas 1758, S\~ao Jos\'e dos Campos, 12227-010, SP, Brazil}

\begin{abstract}
Neutron stars (NS) offer an exceptional opportunity to investigate gravitational physics in extreme environments. In the context of alternative theories of General Relativity (GR), a scalar field non-minimally coupled to gravity can lead to "hairy" NS solutions with a non-zero scalar charge. As a result, binary systems containing at least one "hairy" NS can serve as probes for testing GR through observations. In this work, we analyze the gravitational wave signals from four neutron-star-black-hole (NS-BH) merger events: GW190814, GW200115, GW200105, and GW230529 examine the "hairy" NS hypothesis predicted by the luminal Horndeski theories. Our analysis constrains the scalar charge to be on the order of $10^{-3}$ to $10^{-4}$, with the most stringent limit arising from the recent GW230529 event. In addition, we discuss the implications of our results in relation to the scalarized NS solutions found in the literature.
\end{abstract}



\maketitle

\section{Introduction}
\label{sec:introduction}

Since the historic direct observation of gravitational waves (GWs) in 2015 \cite{Abbott_2016}, numerous additional detections of binary mergers of black holes (BHs) and neutron stars (NSs)have greatly expanded the available data on these extreme gravitational events. To date, nearly 100 compact binary coalescence events have been recorded \cite{Abbott_2021, Abbott_2023}. These GW observations provide increasingly robust opportunities to conduct novel and stringent tests of general relativity (GR) in strong-field regimes (for a comprehensive review, see \cite{Berti_2015}).

GR is widely accepted as the standard theory of gravitation, having withstood numerous experimental validations in various scenarios \cite{Will_2014,Turyshev_2008}. However, there is also a consensus that GR has limitations and is not a fully comprehensive theory of gravity, prompting the exploration of alternative theories that could extend the reach of GR \cite{Berti_2015,Ishak:2018his,Heisenberg_2019}. Both theoretical and observational considerations suggest that GR may require modification in regimes of strong gravitational fields and at cosmological scales. Observationally, the mechanism behind the late-time accelerated expansion of the Universe remains unresolved, with additional gravitational degrees of freedom posited as one possible explanation \cite{Ishak:2018his,Heisenberg_2019}. Beyond-GR theories offer potential solutions to the Hubble constant tension that persists within the framework of the $\Lambda$ CDM model \cite{DiValentino:2021izs}. Additionally, modified gravity models have been proposed to explain the rapid early expansion (inflation) of the Universe. For further motivation and an overview of modified gravity models in strong gravitational field regimes, see \cite{Berti_2015,yunes2024gws} and references therein. A suite of GR-agnostic tests, incorporating parametric corrections, was conducted using compact binary signals \cite{Abbott_2021_GR_test, theligoscientificcollaboration2021tests, LIGO2016lio, LIGOScientific2020tif}. Tests employed novel theory-independent methods—i.e., general tests that are minimally dependent on theoretical assumptions are presented in \cite{Mehta_2023, Krishnendu_2021, maselli2024blackholespectroscopykerr, Shoom_2023}—while also considering different theoretical perspectives \cite{Sennett_2020, Silva_2023, Evstafyeva_2023, Lyu_2022, Mishra_2024, Mishra_2022, Okounkova_2020, Wong_2022, Quartin_2023, Shao_2017}. To date, analysis of GW data has not revealed deviations from GR.

On the other hand, neutron stars present a remarkable setting for probing gravitational and matter physics under extreme conditions. In particular, the binary NS (BNS) merger event GW170817 \cite{Abbott_2017}, accompanied by an electromagnetic counterpart, demonstrated that the speed of GW closely matches the speed of light, with a negligible relative difference of about $10^{-15}$, placing stringent constraints on various gravitational theories \cite{Kase_2019, Copeland_2019, Creminelli_2017, Ezquiaga_2017}. This same GW event provided an intriguing opportunity to constrain the NS matter equation of state (EOS) through measurements of tidal deformations. Additionally, the detection of an NS-BH merger in the GW200115 event \cite{Abbott_2021_NSBH} further allows testing of strong-field gravity in unique ways, e.g. \cite{Lyu_2022, takeda2024gravitationalwaveconstraintsscalartensorgravity}.

It is established, considering scalar-tensor theories with a non-minimally coupled scalar field, that static, spherically symmetric black holes in a vacuum do not acquire additional scalar "hair" \cite{Hawking:1971vc, Bekenstein:1972ny, Faraoni_2017, Sotiriou_2012}. However, for NSs, the presence of matter introduces interactions between the scalar field and matter via gravity-mediated non-minimal coupling. As a result, the scalar field can develop nontrivial profiles in the vicinity of NSs, altering the background geometry through the scalar-gravity coupling. This non-minimal coupling allows for the formation of "hairy" NS solutions that carry a scalar charge, a feature that does not occur for BHs in such models. GR tests in this regard have been extensively explored in the literature and continue to be an active area of research \cite{Anderson_2019, Zhao_2022, Yagi_2016, Shao_2017}.

Among the most popular alternatives to GR we have the Horndeski framework \cite{Horndeski:1974wa}. Horndeski theory comprises the most general scalar-tensor model that is second order in the equations of motion. This last property is crucial as it ensures the avoidance of Ostrogradsky instabilities~\cite{Woodard:2015zca} which are typically associated with higher-order derivatives and can lead to ghost degrees of freedom. The theory also admits both hairy BH \cite{Chatzifotis:2021pak} and NS \cite{Higashino_2023} solutions. Constraints on the general Horndeski theory were obtained in \cite{Arai_2018, Kase_2019, Kreisch_2018, Hou_2018, Bayarsaikhan_2020, Noller_2019} in the cosmological context. A special case of Horndeski theory, namely the Einstein-dilaton Gauss-Bonnet (EdGB) theory, allows for hairy BH solutions through the presence of a scalar charge, and much work has been done to improve constraints on the Gauss-Bonnet coupling constant \cite{Lyu_2022, sänger2024testsgeneralrelativitygw230529, xie2024bayesiansearchmassivescalar, gao2024constraintseinsteindilationgaussbonnetgravityelectric, wang2023constraining, Nair2019iur, Perkins2021mhb}. 

The speed of GW propagation in Horndeski theory has been tightly constrained by observations of the BNS merger event GW170817 \cite{Kreisch_2018, Kase_2019, Copeland_2019, Creminelli_2017, Ezquiaga_2017}. This has driven interest in studying luminal Horndeski theories, a subclass of Horndeski models where the speed of GWs matches that of light. Corrections to the gravitational waveform induced by luminal Horndeski theories have been derived using a post-Newtonian (PN) theory \cite{Higashino_2023}. Additionally, the scalar field coupling constant in these models has been investigated by Bayesian parameter estimation, employing GW signals from NS-BH merger events \cite{Quartin_2023, Niu_2021, takeda2024gravitationalwaveconstraintsscalartensorgravity}.

In this paper, we conduct a statistical analysis to test alternative theories of gravity using observational data from NS-BH binary events cataloged by the LIGO/VIRGO/KAGRA collaboration (LVK), specifically GW190814, GW200115, GW200105, and GW230529 \cite{Abac_2024}. Our focus is on constraining the properties of "hairy" neutron stars predicted by luminal Horndeski theories. In particular, GW230529 reveals evidence of a lower mass gap in the distribution of compact object masses, between 3 \( M_\odot \) and 5 \( M_\odot \), and its long inspiral signal provides a unique opportunity to test GR in a parameter space previously unexplored by strong field tests. Some GR tests for this event were recently carried out in \cite{sänger2024testsgeneralrelativitygw230529, xie2024bayesiansearchmassivescalar,gao2024constraintseinsteindilationgaussbonnetgravityelectric}.
In this work, we examine the event signal in detail to establish new observational constraints on Horndeski gravity at the astrophysical scale. We find that the scalar charge arising from the interaction between the scalar field and matter is on the order of $10^{-3}$ to $10^{-4}$, which is of the same order of magnitude as that found in similar works \cite{takeda2024gravitationalwaveconstraintsscalartensorgravity}, despite being weaker when compared to constraints put by binary pulsar experiments, which place the bound on the order of $10^{-7}$ to $10^{-8}$ \cite{Nair:2020ggs, Quartin_2023}. This result places new observational constraints on the non-minimal coupling. We also discuss the implications of these limits for the fundamental theory of gravity. Our findings do not reveal significant deviations from GR. The structure of this paper is as follows. In Section \ref{theory}, we outline the theoretical framework. Section \ref{meth_data} describes the data sets used in our analysis and presents the main results along with a discussion of their interpretation. Finally, in Section \ref{sec:final}, we provide concluding remarks and discuss potential directions for future research.

\section{Scalar-tensor theories and the modified waveform}
\label{theory}

In this section, we begin our discussion on modifications to the gravitational waveform introduced by a subclass of Horndeski theories and how they are mapped to the parameterized post-Einsteinian (ppE) framework \cite{Yunes_2009}. The inspiral phase of the waveform $h=h(f)$, based on the ppE phenomenological approach, in the frequency domain, is given by 
\begin{equation}
\label{eq:ppE_model}
h = \mathcal{A}_{\rm GR}\left(1+\alpha_{\rm ppE} u^{a_{\rm ppE}}\right) \exp\left(i\left[\Psi_{\rm GR}+\beta_{\rm ppE} u^{b_{\rm ppE}}\right]\right),
\end{equation}
where \( \mathcal{A}_{\rm GR} = \mathcal{A}_{\rm GR}(f)  \) and \( \Psi_{\rm GR} = \Psi_{\rm GR}(f) \) are the amplitude and phase of the waveform as predicted by GR, respectively, \( u = (\pi \mathcal{M} f)^{1/3} \),
\( \mathcal{M} = m \eta^{3/5} \) is the chirp mass, \( m = m_1 + m_2 \) is the total
mass, \( m_1 \) and \( m_2 \) are the masses of the two binary components,
and \( \eta = m_1 m_2/m^2 \) is the symmetric mass ratio. The parameters with the “ppE” subscripts (\( \alpha_{\rm ppE} \), \( a_{\rm ppE} \), \( \beta_{\rm ppE} \), \( b_{\rm ppE} \)) describe potential deviations from the GR predictions.
  
As is well known, Horndeski theories comprise the most general class of scalar-tensor theories with equations of motion that are second order \cite{Horndeski:1974wa}. In this work, we focus on luminal Horndeski theories, i.e. scalar-tensor metric theories that satisfy the requirement that the speed of GWs is equivalent to the speed of light \cite{Arai_2018, Kreisch_2018, Kase_2019, Higashino_2023}. Its action, in the Jordan frame, is given by
\begin{equation}\label{eq:hornaction}
    \mathcal{S} = \int d^4 x \sqrt{-g}\left[G_2(\phi, X) - G_3(\phi, X) \Box \phi + G_4(\phi)R  \right] + \mathcal{S}_m,
\end{equation}
where $g$ is the determinant of the physical metric $g_{\mu\nu}$, $X = -(1/2)\nabla^\mu \phi \nabla_\mu \phi$ is the kinetic term of the scalar field $\phi$, with $\nabla_\mu$ being the covariant derivative, $\Box=g^{\mu\nu}\nabla_\mu \nabla_\nu$ is the d'Alembertian, and $G_4$ is given by
\begin{equation}
    G_4(\phi) = \frac{M_{\rm Pl}^2}{2}F(\phi),
\end{equation}
where $M_{\rm Pl}$ is the Planck mass and $F(\phi)$ is a dimensionless function of $\phi$. We consider matter fields $\mathcal{S}_m$ minimally coupled to gravity. 

It is known that for regular coupling functions $G_i$, $i=2,3,4$, there are no static and spherically symmetric BH solutions with scalar hairs \cite{Hawking:1971vc, Bekenstein:1972ny, Faraoni_2017, Minamitsuji_2022, Minamitsuji_2022_2}. On the other hand, the non-minimal coupling $G_4(\phi)R$ can provide hairy NS solutions \cite{Higashino_2023}, which makes coalescing NS-BH binaries a testing ground for theories given by the action \ref{eq:hornaction}. In this context, neutron stars are endowed with a scalar charge and can exhibit a variety of phenomena depending on the coupling function $G_4(\phi)$. An example that has already been strongly constrained by pulsar timing analysis \cite{Zhao_2022, Kramer_2021} is the mechanism of ``spontaneous scalarization'' \cite{Damour:1993hw, Damour_1996, Doneva_2024}, that takes place when $F(\phi)$ contains even functions of $\phi$, which is the case of the Damour Esposito-Farèse model \cite{Damour:1993hw}, where $F(\phi)=\exp(-\beta \phi^2/ (2M^2_{\rm Pl})$.

The presence of a scalar field changes the local value of the gravitational coupling, and hence the inertial mass of a self-gravitational body. In the Einstein frame, which can be achieved by the conformal transformation $\hat{g}_{\mu\nu} = F(\phi)g_{\mu\nu}$, where the hat denotes the quantity in the Einstein frame and $\hat{g}_{\mu\nu}$ is an auxiliary metric, the scalar field couples to matter through the metric tensor such that it acquires a scalar charge. The quantity
\begin{equation}
    \hat{\alpha} \equiv \frac{M_{\rm Pl}\hat{m}_{,\phi}}{\hat{m}}\Bigg|_{\phi=\phi_0},
\end{equation}
where $\hat{m}(\phi) = m/\sqrt{F(\phi)}$ is the ADM mass in the Einstein frame and the subscript $,\phi$ denotes $\partial/\partial \phi$, can be shown to be directly related to the scalar charge and is hence considered a more fundamental physical quantity instead of its counterpart in the Jordan frame~\cite{Higashino_2023}.

The non-minimally coupled scalar field gives rise to additional propagation modes, namely the "breathing" $h_b$ and "longitudinal" $h_l$ polarizations \cite{Maggiore_2000, Eardley:1973br}. In particular, the mode $h_l$ arises when the scalar field possesses a non-vanishing mass \cite{Higashino_2023}. 
However, in regimes where the mass of the scalar field is much smaller than the binary's orbital frequency, the longitudinal mode can be ignored. Furthermore, as long as $\hat{\alpha} \ll 1$, both the longitudinal and the breathing modes are suppressed by the traceless-transverse polarizations $h_+$ and $h_\times$. Hence, in this work, we will ignore these extra modes of propagation.

The frequency domain waveform components considering a flat background were obtained in \cite{Higashino_2023} for a NS-BH binary system in relative circular motion. Ignoring terms $\mathcal{O}(\hat{\alpha}^4)$, the waveform components are given by
\begin{align}\label{eq:hplus}
    h_+(f) &= -\mathcal{A}_{\rm GR}(f)\left[ 1 - \frac{5\eta^{2/5}\kappa_0\hat{\alpha}^2}{48u_*^2} \right](1+\cos^2\iota)e^{-i\Psi_+}\,\, , \\
    \label{eq:htimes}
    h_\times(f) &= -\mathcal{A}_{\rm GR}(f)\left[ 1 - \frac{5\eta^{2/5}\kappa_0\hat{\alpha}^2}{48u_*^2}\right](2\cos\iota)e^{-i\Psi_\times}\,\, ,
\end{align}
where
\begin{align}
    G_* &\equiv \frac{1}{16\pi G_4(\phi_0)}\,\, , \\
    \kappa_0 &\equiv \frac{G_4(\phi_0)}{\zeta_0 M^2_{\rm Pl}}\,\, , \\
    \zeta_0 &\equiv G_{2,X} + \frac{3G_{4,\phi}^2}{G_4}\Bigg|_{\phi=\phi_0}\,\, , \\ 
    \label{eq:ustar}
     u_* &\equiv (G_* \mathcal{M} \pi f)^{1/3}\,\, ,
\end{align}
$\iota$ is the angle between the orbital angular momentum and the line of sight, and $\phi_0$ is the asymptotic value of the scalar field. Note that in Eq.~(\ref{eq:ustar}) there is a degeneracy between $\mathcal{M}$ and $G_*$, hence the modification to the gravitational coupling can be absorbed into the chirp mass. The waveform phases $\Psi_+$ and $\Psi_\times$ are given by
\begin{align}\label{eq:phase}
    \Psi_+ = \Psi_\times + \frac{\pi}{2} = \Psi_{\rm GR} - \frac{5\eta^{2/5}\kappa_0 \hat{\alpha}^2}{1792 u^7}\,\, ,
\end{align}
where $\Psi_{\rm GR}$ is the GR phase.

Comparison of Eqs. (\ref{eq:hplus}), (\ref{eq:htimes}) and (\ref{eq:phase}) with (\ref{eq:ppE_model}), the ppE parameters can be written as
\begin{align}
    \alpha_{\rm ppE} &= -\frac{5}{48}\eta^{2/5}\kappa_0 \hat{\alpha}^2, \;\;\; a_{\rm ppE} = -2\,\, , \\
    \beta_{\rm ppE} &= -\frac{5}{1792}\eta^{2/5}\kappa_0 \hat{\alpha}^2, \;\;\; b_{\rm ppE} = -7\,\, ,
\end{align}
and the corrections enter at -1PN order relative to the leading GR term. Although there appear to be two deviation parameters, $\kappa_0$ and $\hat{\alpha}$, they are considered as a single correction due to their coupling. Moreover, $\kappa_0$ can be fixed once the coupling functions $G_i$ are fixed to a specific theory.

\begin{figure}
    \centering
    \includegraphics[scale=0.3]{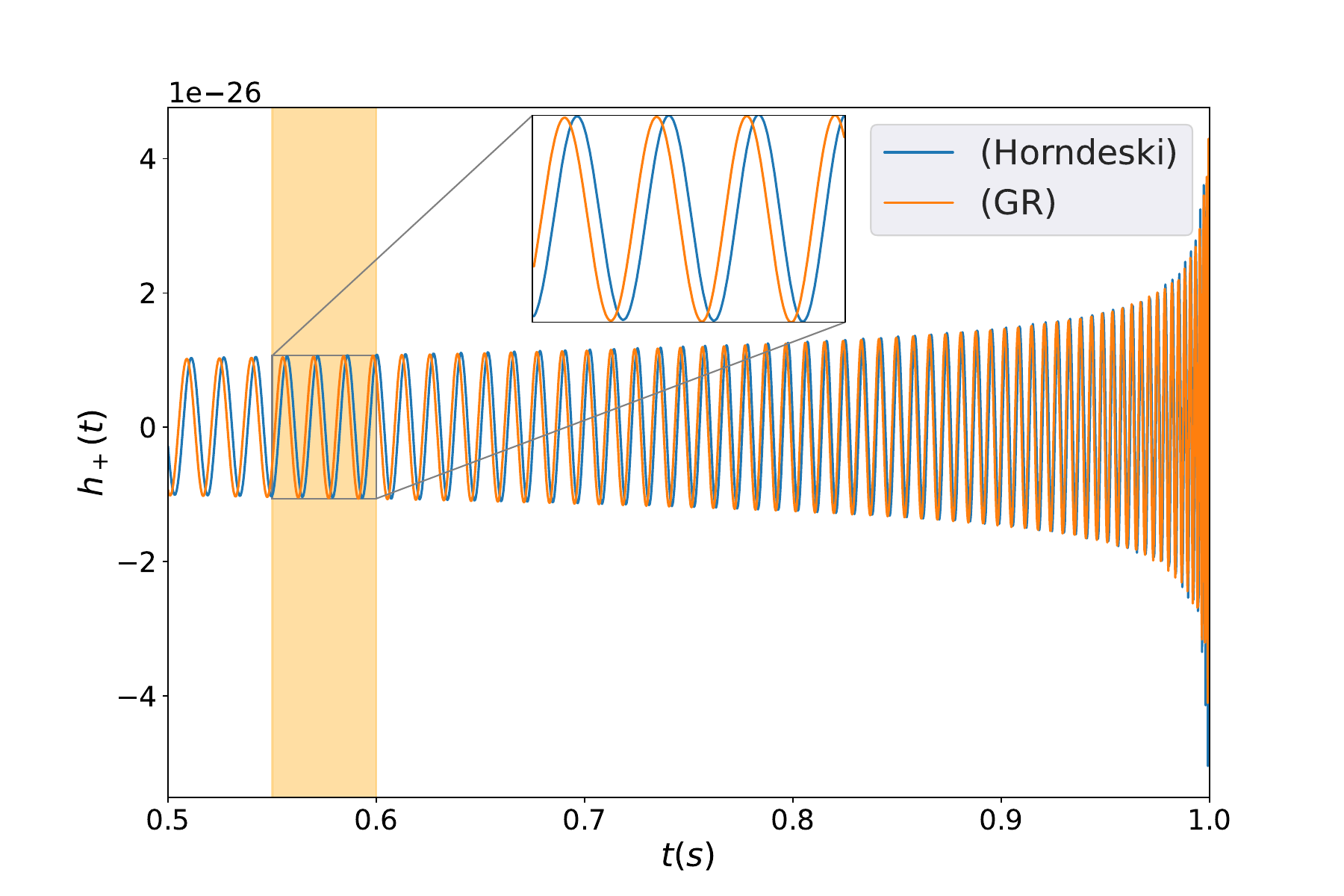}
    \caption{Comparison of simulated polarization signals during the inspiral phase between GR and the luminal Horndeski model. The Horndeski waveform was generated with $\kappa_0 \hat{\alpha}^2 = 0.001$, while standard parameters for both waveforms followed the GW200115 baseline \cite{Abbott_2021}. The simulations were performed using the \texttt{IMRPhenomNSBH} approximant \cite{Thompson_2020}.}
    \label{fig:hcomparison}
\end{figure}

Figure~\ref{fig:hcomparison} compares the simulated polarization signals during the inspiral phase as predicted by GR and the luminal Horndeski models considered in this study. For this comparison, we fixed all binary parameters to the values obtained from the official LIGO analysis (GW200115 baseline), varying only $\kappa_0 \hat{\alpha}^2 \sim 0.001$ to quantify a typical deviation from GR. Even for this small deviation, a significant phase difference emerges, reaching a maximum of approximately $2\pi$ near $\sim 400\ \mathrm{Hz}$. However, the average percentage change in amplitude during the inspiral phase is negligible, at approximately $0.08\%$. Although the amplitude change continues to increase during the merger phase, it is important to note that the ppE model used here is valid only for the inspiral portion of the waveform.

\section{Methodology and Analysis}
\label{meth_data}
Our analysis focuses on the NS-BH coalescence events. In the context of Horndeski theories described by Eq.~(\ref{eq:hornaction}), only the neutron star component can possess a scalar charge. Since we disregard the effects of extra polarization modes beyond those predicted by GR, we include the single-detector events GW200105 and GW230529 in our study. The latter is particularly notable for its extended inspiral signal and the unique nature of one of its components, which lies in the low mass gap.

The gravitational signal from an aligned-spin binary coalescence event depends on four intrinsic parameters that govern the binary's dynamics: the primary and secondary masses, $m_1$ and $m_2$, and the dimensionless spin parameters, $\chi_1$ and $\chi_2$. In our analysis, we introduce an additional intrinsic parameter, $\kappa_0 \hat{\alpha}^2$, which accounts for the possibility of scalar hair in neutron stars. Alongside these intrinsic parameters, there are seven extrinsic parameters related to the detector response: the luminosity distance, $d_L$; the orbital inclination angle, $\iota$; the position of the sky, specified by the right ascension, $\alpha$ and the declination, $\delta$; the polarization angle, $\psi$; the phase at a reference frequency, $\phi_c$; and the coalescence time, $t_c$. Together, these define the full parameter space as follows:

\begin{equation}
\label{eq:parameter_space1}
\theta \equiv\! \Bigl\{m_1, m_2, \chi_1, \chi_2, d_L, \alpha, \delta, \iota, \psi, t_c, \phi_c, \kappa_0 \hat{\alpha}^2\Bigr\}.
\end{equation}

Our analyses are based on Bayesian parameter estimation. For a model $M$ characterized by a set of parameters $\theta$, the posterior probability distribution of $\theta$ given the detector strain data $d$ is expressed as 
\begin{equation}
    p(\theta|d,M) = \frac{\mathcal{L}(d|\theta, M) \pi(\theta|M)}{\mathcal{Z}},
\end{equation}
where $\mathcal{L}(d|\theta,M)$ is the likelihood function, $\pi(\theta|M)$ is the prior and  
\begin{equation}
    \mathcal{Z} = \int \mathcal{L}(d|\theta,M) \pi(\theta|M) \, d\theta
\end{equation}
is the evidence.  

Assuming Gaussian and stationary detector noise, we adopt a standard Gaussian noise likelihood given by  
\begin{equation}
\label{eq:L}
\mathcal{L}(d|\theta,M) \propto \exp{\left[-\frac{1}{2}\sum_I \langle h_I(\theta) - d_I | h_I(\theta) - d_I \rangle \right]},
\end{equation}
where $I$ denotes the detector label. The noise-weighted inner product $\langle \cdot | \cdot \rangle$ is defined as  
\begin{equation}
    \langle a|b \rangle \equiv 4\,\mathrm{Re}\int_{f_{\rm min}}^{f_{\rm max}} \frac{a^*(f) b(f)}{S_{I,n}(f)} \, df,
\end{equation}
where $S_{I,n}(f)$ is the noise power spectral density of detector $I$, obtained from the LVK posterior sample releases \cite{Abac_2024,Abbott_2023}, and $a^*(f)$ is the complex conjugate of $a(f)$.

The minimum frequency, $f_{\rm min}$, is set to $20\ \mathrm{Hz}$ in all analyses, as the detector noise is significant below this threshold. Since the model under consideration is applicable only to the inspiral phase of a coalescing binary, the maximum frequency, $f_{\rm max}$, is chosen as the innermost stable circular orbit (ISCO) frequency for each event. The impact of such a limitation in the signals considered in the analyses is negligible, given the overall low signal-to-noise ratio (SNR) of these events. For non-spinning objects in GR, this frequency is defined as
\begin{equation}
\label{eq:f_ISCO}
    f_{\rm ISCO} = \frac{1}{\pi m \sqrt{6^3}}\,.
\end{equation} 

While the Horndeski theories described by Eq.~\ref{eq:hornaction}) introduce modifications to the ISCO frequency, the deviations from Eq.~(\ref{eq:f_ISCO}) are negligible and can safely be ignored~\cite{takeda2024gravitationalwaveconstraintsscalartensorgravity}.

We use the \texttt{IMRPhenomNSBH} approximant \cite{Thompson_2020} as the GR template for the inspiral phase for all analyses except for the event GW190814, which was analyzed with the \texttt{IMRPhenomXPHM} approximant \cite{Pratten:2020ceb}, since there wasn't a corresponding LVK analysis with the former approximant for this particular event. The \texttt{IMRPhenomNSBH} waveform employs a state-of-the-art phase model derived from the \texttt{NRTidal} approximants \cite{Dietrich:2017aum, Dietrich:2018uni, Dietrich:2019kaq} and an amplitude model parameterized by a single dimensionless tidal deformability parameter. It is designed for aligned-spin systems and is calibrated for mass ratios as low as $q \sim 0.07$, incorporating tidal disruption effects. Additionally, this waveform restricts the primary dimensionless spin to $|\chi_1| < 0.5$, while the secondary spin component is fixed at $\chi_2 = 0$.

Regarding priors, spin priors are constrained by the validity range of the \texttt{IMRPhenomNSBH} approximant: $|\chi_1| < 0.5$ and $\chi_2 = 0$ for the events GW230529, GW200115 and GW200105. For the event GW190814, since the IMRPhenomXPHM supports precessing spins with absolute values up to $0.99$, the spin priors are set accordingly. We assume a uniform prior on $\kappa_0 \hat{\alpha}^2$ over the range $[-0.1, 0.1]$. For all other standard parameters, we adopt the same priors as used in the LVK analysis for each event~\cite{Abbott_2023}. Since the LVK parameter estimation results indicate that the dimensionless spins of the binary components are poorly constrained for the events considered in this study, our specific spin priors are not expected to introduce significant systematic biases.

We use the \texttt{parallel-bilby}\cite{Smith_2020} package and the \texttt{dynesty}\cite{Speagle_2020} sampler to perform Bayesian parameter estimation for the compact binary coalescence events. The sampler settings are configured to balance computational efficiency with the required accuracy. Specifically, we set the number of live points (\texttt{nlive}) to 1000, the number of autocorrelation times (\texttt{nact}) to 10, and employ the acceptance-walk strategy with 100 walks.

For a given parameter $\theta$, we define
\begin{equation}
\label{delta_theta}
\Delta \theta \equiv p(\theta|d,M_{\rm GR}) - p(\theta|d, M_{\rm ppE}),
\end{equation}
which, in practice, quantifies the relative deviation between GR and the ppE model under consideration in this work. 

In this work, we analyze the following events:

\begin{figure*}
    \centering
    \includegraphics[scale=0.3]{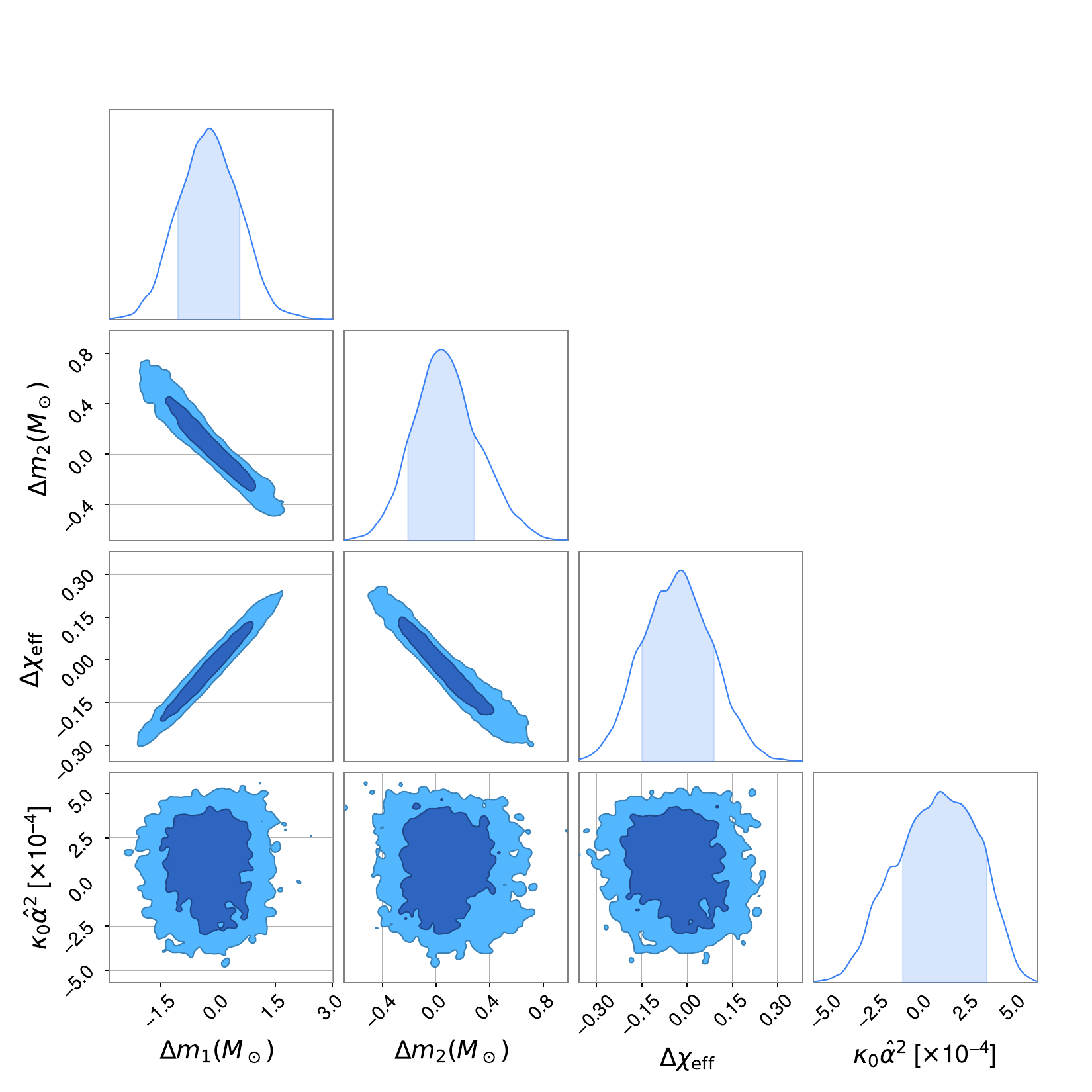} \,\,\,\,\,
    \includegraphics[scale=0.3]{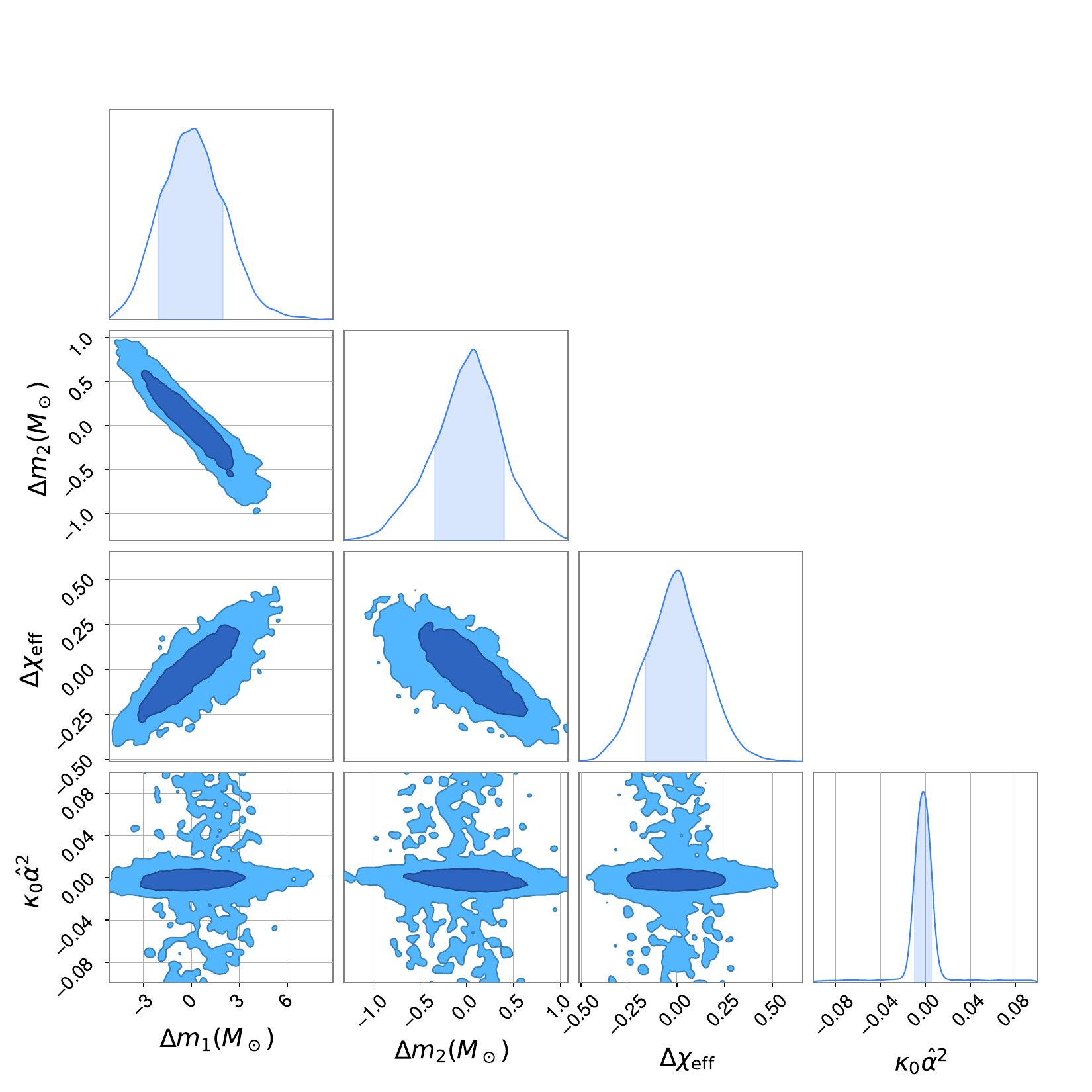}
    \caption{Left panel: Marginalized posterior distributions and 68\% and 95\% confidence level contours of the differences, as defined by Eq.~(\ref{delta_theta}), between the GW230529 posteriors from the ppE model and the LVK posterior sample releases for the source component masses and effective inspiral spin. Right panel: Same as in the left panel, but for the event GW200105.}
    \label{fig:GW230529_corner}
\end{figure*}

\begin{figure*}
    \centering
    \includegraphics[scale=0.3]{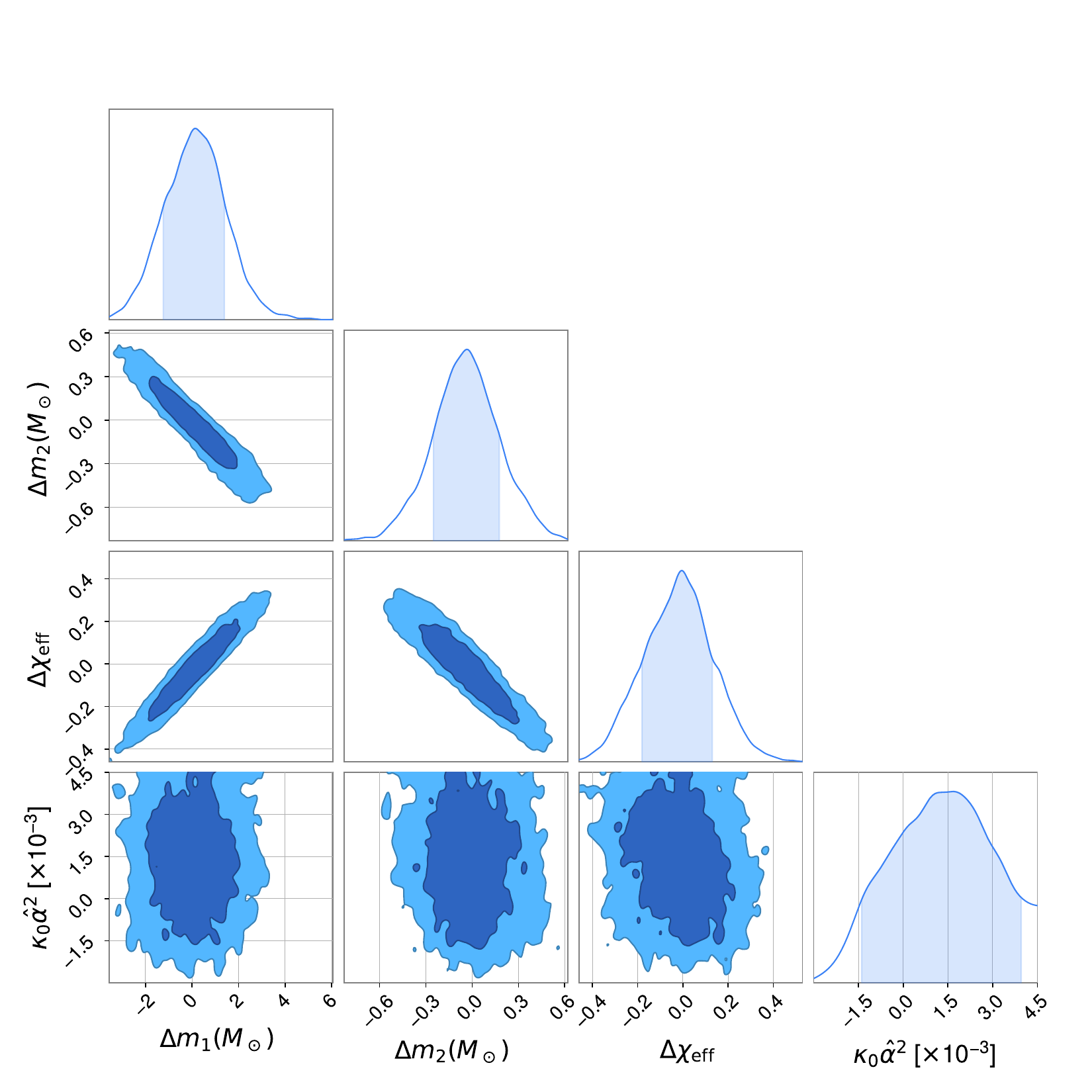} \,\,\,\,\,
    \includegraphics[scale=0.3]{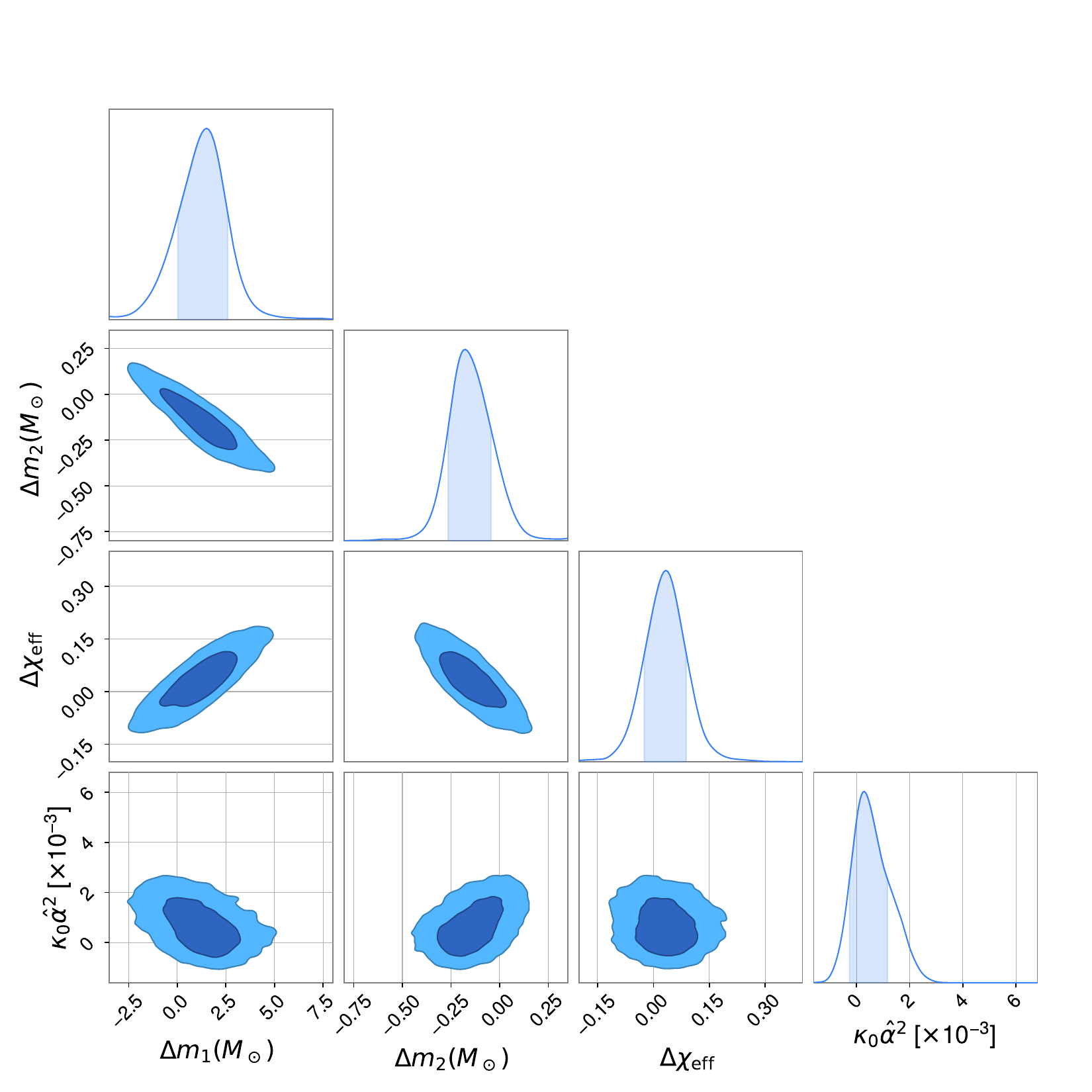}
    \caption{Left panel: Marginalized posterior distributions and 68\% and 95\% confidence level contours of the differences, as defined by Eq.~(\ref{delta_theta}), between the GW200115 posteriors from the ppE model and the LVK posterior sample releases for the source component masses and effective inspiral spin. Right panel: Same as in the left panel, but for the event GW190814.}
    \label{fig:GW200115_corner}
\end{figure*}

\begin{itemize}
    \item \textbf{GW230529} \cite{Abac_2024}: Shortly after the LVK began its fourth observing run (O4), LIGO Livingston detected a potential NS–BH merger, designated GW230529. One of its components had a mass within the lower mass gap. The event was detected with a SNR of approximately 11.6 and a false alarm rate (FAR) of $10^{-3} \, \rm{yr}^{-1}$. The primary and secondary masses were estimated to be $m_1 = 3.6^{+0.8}_{-1.2} \, M_\odot$ and $m_2 = 1.4^{+0.6}_{-0.2} \, M_\odot$ with 90\% confidence level (CL). Although identifying the nature of the components was challenging, LVK's probabilistic classification assigned $\lesssim 3\%$ probability for the primary being a neutron star and $\gtrsim 90\%$ for the secondary component being a neutron star.

    \item \textbf{GW200105} \cite{Abbott_2021_NSBH}: The event GW200105 was a binary coalescence detected by LIGO Livingston with an SNR of 13.9 and a FAR of $2.8 \times 10^{-1} \, \rm{yr}^{-1}$. Despite its relatively high FAR, it is assumed to be of astrophysical origin. The component masses were determined to be $m_1 = 8.9^{+1.1}_{-1.3}\, M_\odot$ and $m_2 = 1.9^{+0.2}_{-0.2}\, M_\odot$. As the primary mass exceeds the maximum mass expected for a neutron star, it was identified as a black hole (BH), while the secondary component was identified as a neutron star with $\sim 96\%$ probability.

    \item \textbf{GW200115} \cite{Abbott_2021_NSBH}: GW200115 was a binary coalescence detected by LIGO Livingston, LIGO Hanford, and Virgo, with a network SNR of 11.3. The primary and secondary masses were estimated as $m_1=5.7^{+1.8}_{-2.1}\, M_\odot$ and $m_2=1.5^{+0.7}_{-0.3}\, M_\odot$, respectively. The secondary mass is consistent with the maximum mass for neutron stars, and there is a $\sim 30\%$ probability that the primary mass falls within the lower mass gap. This event is considered a potential NS–BH merger.

    \item \textbf{GW190814} \cite{Abbott_2020}: GW190814 was detected during the third observing run (O3) by LIGO Livingston, LIGO Hanford, and Virgo with a network SNR of approximately 25. Its unusually high SNR and the number of observed inspiral cycles allowed tight constraints on the component masses: $m_1 = 23.2^{+1.1}_{-1.0}\,M_\odot$ and $m_2 = 2.59^{+0.08}_{-0.09} \, M_\odot$. The nature of the secondary component is uncertain, as its mass exceeds theoretical and observational limits for the maximum neutron star mass. Nevertheless, the possibility of it being a neutron star is not excluded, and it is treated as such in this analysis.

\end{itemize}

Figures \ref{fig:GW230529_corner} and \ref{fig:GW200115_corner} show the relative deviations for the mass of the source component and the effective inspiral spin. The figures also include the posterior distribution for the additional parameter $\kappa_0 \hat{\alpha}^2$. GR samples were obtained from the LVK posterior sample releases \cite{Abbott_2023, Abac_2024}. All constraints on $\kappa_0 \hat{\alpha}^2$ are consistent with GR. In other words, our results are fully compatible with the null hypothesis, $\kappa_0 \hat{\alpha}^2 = 0$. 

In the analysis of GW230529, we set $f_{\rm max} = f_{\rm ISCO} \approx 830$ Hz and analyzed 128 seconds of data. For this case, we find
\begin{equation}\label{eq:alphaGW230529}
    \kappa_0 \hat{\alpha}^2 = (1.0^{+2.5}_{-2.0}) \times 10^{-4} \,\, ,
\end{equation}
with 90\% CL. As shown in the left panel of Fig.\ref{fig:GW230529_corner}, the 2D plots indicate that there is no correlation between the extra parameter and the standard parameters displayed. The relative deviations for the masses of the source component and the effective inspiral spin are $\Delta m_1 = -0.023^{+0.80}_{-0.82}\, M_\odot$, $\Delta m_2 = 0.04^{+0.25}_{-0.25}\, M_\odot$, and $\Delta \chi_{\rm eff} = -0.02^{+0.11}_{-0.13}$, respectively. In general, when examining $\Delta \theta$ for all baselines of the model, all parameters $\Delta \theta$ are compatible with $\Delta \theta = 0$ at the 1$\sigma$ CL.

For the analysis of GW200105, we set $f_{\rm max} = f_{\rm ISCO} \approx 402 \, \rm{Hz}$. For this event, we find
\begin{equation}\label{eq:alphaGW200105}
    \kappa_0 \hat{\alpha}^2 = (-0.9^{+6.2}_{-8.6}) \times 10^{-3},
\end{equation}
with 90\% CL. The relative deviations for the other standard parameters are $\Delta m_1 = 0.02^{+1.8}_{-2.2}\, M_\odot$, $\Delta m_2 = 0.08^{+0.32}_{-0.42}\, M_\odot$, and $\Delta \chi_{\rm eff} = 0.01^{+0.14}_{-0.18}$. Again, no significant deviations from the GR are noted. Furthermore, for this signal, there is no correlation between the extra parameter and the intrinsic parameters of the binary system.

For the event GW200115, we set $f_{\rm max} = f_{\rm ISCO} \approx 604 \, \rm{Hz}$ and analyzed 32 seconds of data. We find
\begin{equation}\label{eq:alphaGW200115}
    \kappa_0 \hat{\alpha}^2 = (1.1^{+2.9}_{-2.5}) \times 10^{-3}.
\end{equation}
As can be seen in the 2D plots shown in Figure \ref{fig:GW200115_corner}, there is no correlation between the additional parameter and the standard parameters displayed. 
Similarly to the previous events, we note that $\Delta \theta_{i} = 0$ at 1$\sigma$ for all baseline parameters.

Finally, for the event GW190814, we set $f_{\rm max} = f_{\rm ISCO} \approx 162 \, \rm{Hz}$ and considered 4 seconds of data. We find
\begin{equation}\label{eq:alphaGW190814}
    \kappa_0 \hat{\alpha}^2 = (2.6^{+9.0}_{-5.3}) \times 10^{-4},
\end{equation}
with 90\% CL. This is the only event to show a correlation between $\kappa_0 \hat{\alpha}^2$ and the source component masses (see the right panel of Fig. \ref{fig:GW200115_corner}), which we believe could be attributed to its higher SNR compared to the other events considered in this work.

It can be seen from the constraints provided by Eqs.~(\ref{eq:alphaGW230529}), (\ref{eq:alphaGW200105}), (\ref{eq:alphaGW200115}), and (\ref{eq:alphaGW190814}) that the results are consistent with the predicted behavior of the correction parameter $-1$ PN, $\delta\phi_{-2}$~\cite{sänger2024testsgeneralrelativitygw230529}, showing no significant deviation from GR. In particular, there is a general preference for positive values of $\kappa_0 \hat{\alpha}^2$, suggesting a tendency towards $\kappa_0 > 0$. However, this preference should be interpreted with caution, as it may reflect systematic errors arising from the limited number of analyzed events and the overall low SNR of the data.

If we set $\kappa_0 \approx 1/2$, as is the case for Brans-Dicke theories or theories of spontaneous scalarization of NSs~\cite{Higashino_2023, Quartin_2023, takeda2024gravitationalwaveconstraintsscalartensorgravity}, we obtain a bound for the absolute value of $\hat{\alpha}$:
\begin{equation}\label{eq:boundalpha}
   | \hat{\alpha}| \leq 0.028,
\end{equation}
which follows from the tighter constraint given by Eq.~(\ref{eq:alphaGW230529}) derived from the analysis of the event GW230529. This constraint is consistent with those found in \cite{takeda2024gravitationalwaveconstraintsscalartensorgravity, Niu_2021, Quartin_2023}, where similar conclusions were drawn with respect to the scalar charge and the additional degrees of freedom in the theories considered.

It is known that the scalar charge and the non-minimal coupling are related to the neutron star (NS) equation of state (EOS). From our statistical inference, discussed above, we can derive some quantitative estimates. Assuming a static, spherically symmetric background and an NS with nearly constant density, the relationship between $\hat{\alpha}$, the non-minimal coupling $F(\phi)$, and the EOS can be approximated as~\cite{Higashino_2023}
\begin{equation}\label{eq:alphaEOS}
    \hat{\alpha} \approx -\frac{\xi_c}{2}(1-3w_c),
\end{equation}
where $w_c = P_c/\rho_c$ is the EOS parameter, with $P_c$ and $\rho_c$ representing the central pressure and density, respectively, and
\begin{equation}
   \xi_c = M_{\rm Pl}\frac{F_{,\phi}}{F}\Bigg|_{\phi=\phi_c},
\end{equation}
where $\phi_c = \phi(r=0)$ is the value of the field at the center of the star. Note that Eq.~(\ref{eq:alphaEOS}) is valid only for $w_c < 1/3$. Typically, $w_c \sim 0.1$ for relativistic stars and $w_c \ll 1$ for non-relativistic stars.

For BD theories, $\xi_c = -2Q$, where $Q$ is the respective coupling constant. The bound on $\hat{\alpha}$ given by Eq.~(\ref{eq:boundalpha}) considering $w_c$ for relativistic stars yields 
\begin{equation}\label{eq:qbound}
    |Q| \lesssim 0.04,
\end{equation}
which is close to the bound found in \cite{takeda2024gravitationalwaveconstraintsscalartensorgravity} ($|Q| \leq 0.055$), which was obtained by considering the ``breathing'' polarization mode of the scalar field. However, it should be noted that slight variations in the EOS parameter induce order-of-magnitude variations in $|Q|$, e.g., by choosing $w_c = 0.2$, we get $|Q| \lesssim 0.1$. Hence, further analysis incorporating a choice of EOS is required. Nevertheless, the bound given by Eq.~(\ref{eq:qbound}) is one order of magnitude higher than the tightest constraints on this parameter given by solar system tests ($|Q| \leq 2.5 \times 10^{-3}$) \cite{Will_2014}. This suggests that future detectors with higher sensitivities may place even tighter constraints than those achieved by solar system tests.

According to the relation $|Q| = \left[2(3+2\omega_{\rm BD})\right]^{-1/2}$~\cite{De_Felice_2010}, where $\omega_{\rm BD}$ is the BD parameter, the bound given by Eq.~(\ref{eq:qbound}) translates to
\begin{equation}\label{eq:omegabd}
    \omega_{\rm BD} \gtrsim 155,
\end{equation}
which is consistent with the strongest constraints found to date in the context of GW analysis ($\omega_{\rm BD} > 110.55)$ \cite{Tan_2024}, where they analysed both GW190814 and GW200115 considering higher waveform harmonics. The previous bounds obtained through the analysis of the GW data yield $\omega_{\rm BD} \gtrsim 40$ \cite{Niu_2021} and $\omega_{\rm BD} \gtrsim 81$ \cite{takeda2024gravitationalwaveconstraintsscalartensorgravity}. Constraints on the BD parameter are much stronger in the context of cosmological analysis using Cosmic Microwave Background (CMB) data, with bounds of $\omega_{\rm BD} > 890$ \cite{Avilez_2014}. The strongest bound on this parameter to date ($\omega_{\rm BD} > 40000$) was put by measurements from the Cassini mission~\cite{Bertotti:2003rm}.
 
Although $\hat{\alpha}$ has been shown to decrease with increasing component mass for a fixed value of $\kappa_0$ and the non-minimal coupling~\cite{takeda2024gravitationalwaveconstraintsscalartensorgravity}, our findings do not exhibit this trend. In particular, our most stringent constraint on $\kappa_0 \hat{\alpha}^2$ arises from the analysis of GW230529, which involves the lowest putative NS mass among all the events considered here. Assuming that GW190814 is an NS-BH merger event, it should yield the lowest estimate for the ppE parameter. However, this effect may be attributed to either low SNR or systematic bias arising from the aforementioned assumption.

\section{Conclusion}
\label{sec:final}

In this work, we have analyzed GW signals from four NS-BH merger events: GW230529, GW200105, GW200115, and GW190814, to probe the possible presence of scalarized NS predicted by luminal Horndeski theories. Our analysis focused on extracting constraints on the scalar charge, a fundamental parameter associated with the non-minimal coupling of a scalar field to gravity. For each event, we conducted a robust Markov-Chain Monte-Carlo analysis examination using the ppE model to incorporate possible deviations from GR and analyzed the posterior distributions of the additional parameters, including the scalar charge. In all cases, we found that the scalar charge remained consistent with GR, with upper limits on the parameter typically of the order of $10^{-4}$, and no significant deviation from GR was observed. In particular, the most stringent constraints came from the GW230529 event, with $\kappa_0 \hat{\alpha}^2$ at the level of $(1.0^{+2.5}_{-2.0}) \times 10^{-4}$. Despite the large uncertainties in some cases, especially those with a lower SNR, the results of GW190814 suggest a potential mild correlation between the extra parameter $\kappa_0 \hat{\alpha}^2$ and the source component masses of the binary system, although this was not observed for other events. These findings highlight the importance of the SNR in constraining scalar charge parameters, and further studies with higher precision measurements will be crucial in tightening these bounds.

In terms of physical implications, the absence of significant deviations from GR in our analysis indicates that scalarized NS solutions, although theoretically possible, are not strongly favored by the current data from these NS-BH mergers. The results are fully compatible with the null hypothesis, confirming that GR remains a robust description of gravitational interactions in the strong-field regime probed by compact binary mergers.

The upcoming observational runs of LIGO, Virgo, and KAGRA are expected to provide more precise measurements of compact binary mergers, which will allow for tighter constraints on deviations from GR, including potential scalar field interactions. Events with higher SNR, particularly those from nearby sources or future detections, will yield stronger limits on the scalar charge. Furthermore, the development of more advanced models for the dynamics of scalar fields in compact objects will be crucial for interpreting future data and refining our understanding of gravitational alternatives to GR. Studies using real data that account for the stability of scalarized solutions, the equation of state of neutron star matter, and the potential for gravitational wave signatures from such systems will continue to enhance our understanding of gravity. We hope to present further results on this topic in future communications.
\\

\begin{acknowledgments}
\bigskip
\noindent The authors thank the referees for their valuable comments and suggestions, which have helped improve the clarity and significance of the results presented in this work. R.M.S. thanks CNPq (141351/2023-3) for financial support. R.C.N thanks the financial support from the CNPq under the project No. 304306/2022-3, and the FAPERGS under the project No. 23/2551-0000848-3. 
\end{acknowledgments}

\bibliographystyle{apsrev4-1}
%

\end{document}